\documentclass[a4paper]{jpconf}
\usepackage{graphicx}
\begin{document}
\title{Dynamics of dinuclear system formation and its decay in heavy ion collisions}

\author{Avazbek Nasirov$^{1,2}$, Giorgio Giardina$^{3}$, Giuseppe
Mandarglio$^{3}$, Marina Manganaro$^{3}$, Werner Scheid$^{4}$}

\address{$^1$Joint Institute for Nuclear Research, 141980 Dubna, Russia}
\address{$^2$Institute of Nuclear Physics, 100214, Tashkent, Uzbekistan}
\address{$^3$Dipartimento di Fisica dell' Universit\'a di Messina, 98166 Messina,
  and Istituto Nazionale di Fisica Nucleare, Sezione di Catania, Italy}
\address{$^4$Institute f\"ur Theoretische Physik der Justus-Liebig-Universit\"at, Giessen, Germany}
\ead{nasirov@jinr.ru}

\begin{abstract}
A variety of phenomena connected with the formation
of a dinuclear complex is observed in the heavy ion collisions at low energies.
 The dinuclear system model allows us to analyze
the experimental data and to interpret them by comparison of
 the partial capture, fusion and evaporation residue cross sections
 measured for the different reactions leading to the same compound nucleus.
 The comparison of theoretical and experimental values of
 the mass and angular distributions of the reaction products
 gives us a detailed information about reaction mechanism forming
 the observed yields.
 The observed very small cross sections of the evaporation residues may be
explained by the strong fusion hindrance and/or instability of the heated
and rotating compound nucleus and smallness of its survival probability.
The fusion hindrance arises due to competition between complete fusion
and quasifission while the smallness of survival probability is connected
with the decrease of the fission barrier at large excitation energy and angular
momentum of compound nucleus.
\end{abstract}

\section{Introduction}

In the heavy ion collisions at low energies, a variety of phenomena is
observed, which are connected with the formation of a dinuclear complex.
Deeply inelastic collisions have been studied extensively over a wide range
of energies and masses \cite{Wilpert}. The  multinucleon
transfer and energy dissipation mechanisms of these reactions are
very similar with the ones of the full momentum transfer (capture) reactions.
The difference between
deep inelastic collisions and quasifission reactions is determined by the
lifetimes of the dinuclear system (DNS) formed as intermediate system
which consists of interacting nuclei with indestructible corns and excited nucleons in
high-lying quantum states. Some of these excited nucleons form the neck or overlap
region between constituents of DNS. The nucleon exchange mechanism
at dissipation of the relative motion energy of the colliding nuclei
was theoretically studied in Ref.\cite{Isotope,Sharing}.
The basic points of those methods were developed to study deep inelastic and
quasifission processes which are used to explain the hindrance to complete fusion
\cite{GiarSHE}.

 The set of the successful experimental results in the synthesis of superheavy elements
 stimulated great efforts to experimental and theoretical investigations
 of the fusion-fission processes with massive nuclei. The very small cross sections
$\sigma_{ER}$  of the evaporation residues observed in the recent
 experiments \cite{Oganessian,SHofmann,Morita} may be explained by the strong fusion
hindrance and/or instability of the heated and rotating compound nucleus \cite{FazioPRC2005}.

The experimental knowledge about fusion-fission reactions  at sub- and
near-barrier energies has grown considerably in the last twenty years.
   The theoretical models are able to reproduce  the main features
of such processes, even to make  predictions of the cross sections for synthesis of
superheavy elements which are  more or less  close to the experimental data.
But properly understanding the
fusion dynamics for heavy systems requires many more ingredients. The necessity of
more experimental data to disentangle various concurrent effects, is clearly felt.
A full understanding of all steps of the reaction dynamics is very important for
the challenging issue of superheavy element production and new isotopes far from
the valley of stability.

Dynamics of complete fusion and role of the entrance  channel in formation of the
reaction products in heavy ion collisions  are questionable or they have different
interpretation still nowadays.
 For example, what mechanism of fusion makes the main contribution  to formation of
compound nucleus: an increase of the neck between interacting nucleus or
multinucleon transfer at a relatively restricted neck size?
How large is  the overlap between angular momentum distributions of dinuclear system
and compound nucleus which determine the angular distribution of  reaction products,
cross sections of evaporation residue, fusion-fission and quasifission products?
There is an ambiguity in separation of fusion-fission fragments from the quasifission
and fast fission products. Still unclear surely a law of the distribution of the
excitation energy of DNS between its constituent fragments. Therefore,
it is interesting to study the mechanism of deep inelastic  and  quasifission reactions
forming binary fragments accompanied by neutron and gamma quantum emissions.
The analysis of a correlation angular, mass-energy distributions of the registered fragments and
neutron and gamma quantum characteristics allows us to obtain useful knowledge
about the complete fusion mechanism.

\section{Difference and similarity of  quasifission and deep inelastic reactions
}
\label{Section2}

An investigation of dissipation dynamics of the deep inelastic and quasifission reactions
is useful to establish relaxation times of different degrees of freedom
in heavy ion collisions at low energies. From an analysis of the experimental data of
the mass-angle distributions of
the binary reaction products and gamma-quanta of giant dipole resonances
we can conclude that a study of the DNS stage of these dissipative processes
is perspective. Because a sufficient part of the reaction time of the
dissipative processes in heavy ion collisions belongs to this stage,
particularly for the massive system, as well as for the case of colliding nuclei
with nearly equal charge (mass) numbers.
Qualitative difference between deep inelastic and quasifission reactions is
that the full momentum transfer does not take place in the former mechanism
while it occurs in the latter one. The classification of the experimental data
of binary products as ones corresponding to the deep inelastic collision or
capture reactions (quasifission and fusion-fission) is based on the mass distribution
characteristics only: projectile-like and target-like products with the total
kinetic energies (TKE)  around the Viola systematics \cite{Viola} are considered
as products of deep inelastic collisions. The authors of Ref. \cite{KellerPRC36}
attributed all events with energy losses larger than 8--15 MeV and with fragments
of below 40 mass units to deep inelastic collision at the analysis of the
experimental data of the $^{32}$S+$^{182}$W reaction.

\begin{figure}[t]  
\begin{center}
\includegraphics[width=28pc]{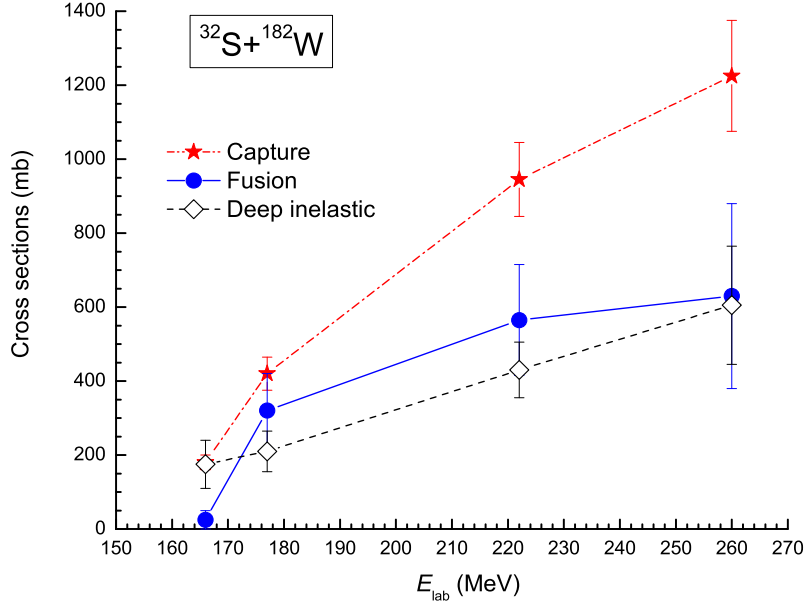}
\vspace*{-0.5cm}
\caption{\label{CaptDICFus} The experimental data for capture (dot-dashed line),
complete fusion (solid line)  and deep inelastic collision (dashed line) events
 for the $^{32}$S+$^{182}$W reaction \cite{KellerPRC36}.}
\end{center}
\end{figure}

 The capture, deep inelastic and complete fusion reaction
cross sections are presented in Fig. \ref{CaptDICFus}. The difference between
capture and complete fusion cross sections is the quasifission cross section:
\begin{equation}
\sigma_{\rm cap}=\sigma_{\rm fus}+\sigma_{\rm qfis}.
\end{equation}
Unfortunately, a possibility of the mixing mass-angle distributions
of the quasifission and deep inelastic reactions is not studied well
because the nature of quasifission has been not established yet.

\begin{figure}[t]  
\begin{center}
\includegraphics[width=28pc]{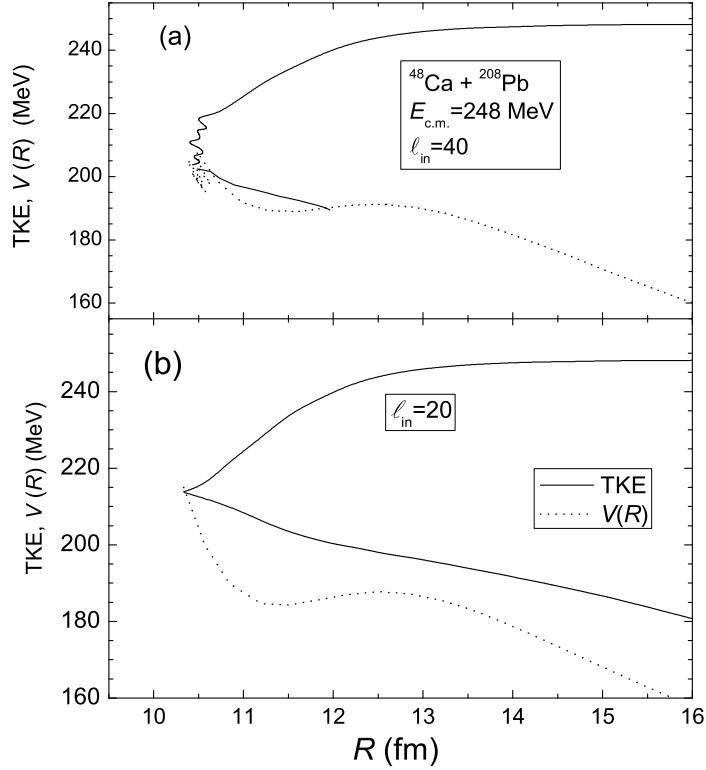}
\vspace*{-6.0cm}
\caption{\label{CapDIC} Capture (a) and deep inelastic collision (b) for the
$^{48}$Ca+$^{208}$Pb reaction.}
\end{center}
\end{figure}

The difference between capture and deep inelastic reactions can be found
theoretically from the analysis of the results of dynamical calculations by solving
the equations of motion for the radial distance (Eq. \ref{EqR}),
angular momentum (Eq. \ref{EqL}) and surface vibrations for quadrupole
and octupole multipolarity (Eq. \ref{EqTeta}) (see Fig. \ref{CapDIC})  for the
$^{48}$Ca+$^{208}$Pb reaction.

The quasifission occurs at the full momentum transfer and
it belongs to the capture reactions when the system traps into
potential well as in Fig. \ref{CapDIC}(a) and the decay time of DNS, which is
formed in this mechanism, is determined by the size of the potential well,
by the values of excitation energy $E^*_{\rm DNS}$ and friction coefficients
(radial and tangential).

In the deep inelastic reactions, the DNS is not trapped into the potential well
because the DNS momentum decreases up to zero and the relative distance between
centers of mass of projectile and target nuclei reaches the minimum value
(see Fig. \ref{CapDIC}(b). Then the relative distance
$R$ increases due to repulsive forces. Although the kinetic
energy of the relative motion goes on to be dissipated the DNS can overcome
the Coulomb barrier from intrinsic part to outside of the potential well
and we observe two products after interaction and exchanging nucleons.
Therefore, the interaction time of colliding nuclei in deep inelastic
collision is small in comparison with the one in capture reactions.
The friction coefficients and the size of the potential well play a crucial
role in the calculations of capture events by solving the following equations
\cite{GiarSHE,NasirovRauis}:
\begin{eqnarray}
 \label{EqR} \mu(R,\alpha_1,\alpha_2)\ddot R +
 \gamma_{R}(R,\alpha_1,\alpha_2)\dot R(t)= F(R),\hspace{2.0cm}\\
 \label{maineq2} F(R,\alpha_1,\alpha_2)=
 -\frac {\partial V(R,\alpha_1,\alpha_2)}{\partial R}-
 \dot R \hspace{-0.3mm}^2 \frac {\partial \mu(R)}{\partial R}\,,\hspace{2.5cm}\\
 \label{EqTeta} D_{\beta_i}\ddot \beta_i(t)+
 \gamma_{\beta}(R)\dot \beta_i(\alpha_1,\alpha_2,t)= F_{\beta_i}(R)\,\hspace{2.0cm}\\
 \label{FTeta} F_{\beta}(R)=
 -\frac {\partial V(\beta_i,\alpha_1,\alpha_2)}{\partial \beta_i}\hspace{2.5cm}\\
 \label{EqL}\frac{dL}{dt}=\gamma_{\theta}(R,\alpha_1,\alpha_2)R(t)\left(\dot{\theta}
 R(t) -\dot{\theta_1} R_{1eff} -\dot{\theta_2} R_{2eff}\right), \\
 L_0=J_R(R,\alpha_1,\alpha_2) \dot{\theta}+J_1 \dot{\theta_1}+J_2 \dot{\theta_2}\,,\hspace{3.35cm} \\
 E_{\rm rot}=\mu(R,\alpha_1,\alpha_2){\dot R}^2/2+\frac{J_R(R,\alpha_1,\alpha_2) \dot{\theta_{}}{}^2}2+\frac{J_1
 \dot{\theta_1}^2}2+\frac{J_2 \dot{\theta_2}^2}2\,,\hspace{2.45cm}
 \end{eqnarray}
 where $R\equiv R(t)$ is the relative motion coordinate; $\dot R(t)$
 is the corresponding velocity; $\alpha_1$ and $\alpha_2$ are the orientation
 angles between beam direction and axial symmetry axis of the projectile
 and target, respectively; $L_0$ ($L_0=\ell_0 \hbar$) and $E_{\rm rot}$ are defined by
 initial conditions; $J_R$ and $\dot\theta$, $J_1$ and
 $\dot\theta_1$, $J_2$ and $\dot\theta_2$ are  moment of inertia
 and  angular velocities of the DNS and its fragments, respectively;
$\gamma_R$ and $\gamma_{\theta}$ are the friction coefficients for the relative
 motion along $R$ and the tangential motion when two nuclei roll on
 each other's surfaces, respectively; $\omega_{\lambda}^{(i)}$,
$\gamma_{\lambda}$ and $D_{\lambda}^{(i)}$
are frequency, damping and mass coefficients for the surface vibrations
with multipolarity
 $\lambda$, respectively;
 $V(R,\alpha_1,\alpha_2)$ is the  nucleus-nucleus
 potential calculated by the double folding procedure \cite{GiarSHE,FazioJPSJ72}.

 The capture includes complete fusion and quasifission events.
 The quasifission time is estimated by the formula
\begin{equation}
\tau_{\rm DNS}(T_Z)=\frac{\hbar}{\Gamma_{\rm qfiss}(T_Z)}
\end{equation}
if we know the excitation energy  $E^*_{\rm DNS}$ and quasifission
barrier $B_{\rm qf}$ of the dinuclear system for its decay in
fragments with charge numbers $Z$ and $Z_{\rm tot}-Z$, by using the
one-dimensional Kramers rate \cite{Kramers,Grange,Froeb92}
\begin{eqnarray}
\Gamma_{\rm qfiss}(\Theta)&=&K^{(\rm DNS)}_{\rm rot}/K^{(\rm CN)}_{\rm rot} \,\omega_m
\left(\sqrt{\gamma^2/(2\mu_{\rm qf})^2+\omega^2_{\rm qf}}-\gamma/(2\mu_{\rm qf})
\right)\nonumber\\
&\times&\exp\left(-B_{\rm qf}/T_Z)\right)/(2\pi\omega_{\rm qf}).
\end{eqnarray}
\begin{figure}[t]  
\begin{center}
\includegraphics[width=28pc]{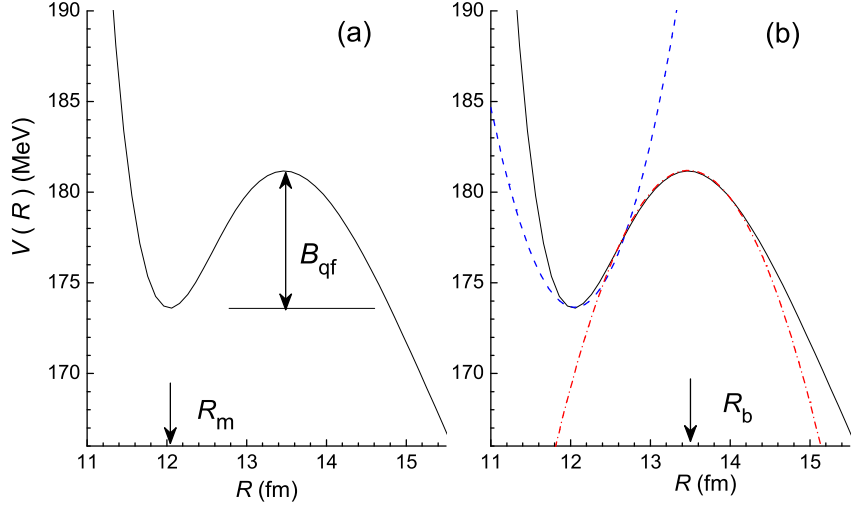}
\vspace*{-0.75cm} \caption{\label{Omegas} (a) Quasifission barrier
is the depth of the well in the nucleus-nucleus interaction
potential $V(R)$. (b) The potential well is replaced by the
harmonic oscillator with frequency $\omega_{\rm m}$ and  the
potential barrier is replaced by the inverted harmonic oscillator
with frequency $\omega_{\rm b}$ which are used to calculate the
decay time of DNS into two fragments.}
\end{center}
\end{figure}
Here the frequencies $\omega_{\rm m}$ and $\omega_{\rm b}$ are found by the
harmonic oscillator approximation to the shape of the nucleus-nucleus potential
$V(R)$  for a given DNS configuration $(Z,Z_{\rm tot}-Z)$ on
the bottom of its pocket placed at $R_{\rm m}$ and  on the top
(quasifission barrier) placed at $R_{\rm b}$ (see Fig.
\ref{Omegas}), respectively:
\begin{eqnarray}
 \omega_m^2&=&\mu_{\rm qf}^{-1}\left|\frac{\partial^2 V(R)}{\partial
R^2}\right|_{R=R_m}\,,\\
\omega_{\rm b}^2&=&\mu_{\rm qf}^{-1}\left|\frac{\partial^2 V(R)}{\partial
R^2}\right|_{R=R_{\rm b}}.
\end{eqnarray}
The nucleus-nucleus potential
$V(Z,A,\ell,R)$ includes the Coulomb $V_{\rm Coul}(Z,A,R)$,
nuclear $V_{\rm N}(Z,A,R)$ and rotational $V_{\rm
rot}(Z,A,R,\ell)$ parts:
\begin{equation}
\label{Vpot} V(Z,A,\ell,R)=V_{\rm Coul}(Z,A,R)+V_{\rm
N}(Z,A,R)+V_{\rm rot}(Z,A,R,\ell),
\end{equation}
where $R$ is the distance between the centers of the nuclei. Details
of the calculation can be found in Refs. \cite{FazioEPJ2004,Nasirov05}.

The calculated values of $\hbar\omega_{\rm m}$ and $\hbar\omega_{\rm qf}$ were
equal to 46.52 MeV and 22.37 MeV, respectively. The used value of the friction
coefficient $\gamma$ is equal to $8\cdot 10^{-22}$ MeV fm$^{-2}$s
which was found from our calculations;
$\mu_{\rm qf}\approx\mu=A_1\cdot A_2/A_{\rm CN}$, where  $A_1$ and $A_2$
are the mass numbers of the quasifission fragments.

 The collective enhancement factor $K_{\rm rot}$ of
the rotational motion  to the level density  should be
included because the dinuclear system is a good rotator. It is
calculated by the well known expression \cite{Junghans}:
 \[K_{\rm rot}(E_{\rm DNS}) =
 \left\{\begin{array}{ll}(\sigma_{\bot}^2-1)f(E_{\rm DNS})+1,  \hspace*{0.2 cm}  \rm{if}
\ \  \sigma_{\bot}>1 \ \  \
  \\ 1,  \hspace*{0.2cm}  \rm{if} \ \
  \it \sigma_{\bot}\le \rm 1\:,
 \end{array}
 \right.
 \]
where  $\sigma^2_{\bot}=J^{(\rm DNS)}_{\bot} T/\hbar^2$;
$f(E)=(1+\exp[(E-E_{\rm cr})/d_{\rm cr}])$;  $E_{\rm cr}=120
\widetilde{\beta}_2^2 A^{1/3}$ MeV; $d_{\rm cr}=1400
\widetilde{\beta}_2^2 A^{2/3}$. $\widetilde{\beta}$ is the
effective quadrupole deformation for DNS. We find it from the
value of the DNS moment of inertia relative to rotation around
axis perpendicular to the line connecting the mass centers of
nuclei which is calculated by the formula:
\begin{eqnarray}\label{jdns2}
  \hspace{0.6 cm} \mathcal{J}^{DNS}_{\bot}=\mathcal{J}_1+\mathcal{J}_2+
  M_1 d^{(1)2}_{\bot}+M_2 d^{(2)2}_{\bot},
\end{eqnarray}
 where $d^{(i)}_{\bot}$   is the distance between the
 center of mass of the fragment $i$ ($i=1,2$) and the line
 corresponding to the largest  moment of inertia of DNS (see Fig. 11 in Ref.\cite{AnisEPJA34}.

The DNS can have  a  long lifetime ($\tau_{\rm DNS}$) if it is
formed at capture: the value of $\tau_{\rm DNS}$ depends on the
depth of the potential well ($B_{\rm qf}$), DNS excitation energy
($E^*_{\rm DNS}$) at the given value of angular momentum $\ell$
and its moment of inertia $\mathcal{J}^{DNS}_{\bot}$
\cite{AnisEPJA34}.

Due to the shell effects as the quantum states of the
neutron and proton systems of the interacting nuclei we have the DNS as a molecule  of nuclei
which does not fuse immediately. Therefore, the peculiarities of
the DNS stage influences on formation of the reaction products
 with magic proton or neutron numbers.  Shell effects are observed as cluster
states in the large amplitude collective motions of nuclei. The observed
cluster emission, mass-charge distribution of the quasifission fragments
and spontaneous asymmetric fission of Th, U and Cf  isotopes proved the strong
role of shell structure. Reactions of heavy ion collisions and fission
(spontaneous and induced) processes can be studied well using the DNS concept.

A characteristic feature of the deep inelastic processes is that
the target and projectile identity is largely preserved. In Ref. \cite{RudolfNPA330},
 only for low  TKE of the outgoing products (energies below the Coulomb repulsion,
$V_{\rm C}$, of touching spheres) a small mass drift is found.
In these studies, it was difficult to establish
the magnitude of such a drift unambiguously, because a relatively small initial mass
asymmetry made it difficult to separate drifting projectile- and target-like
distributions from the wings of a mass symmetric component.

\begin{figure}[t]  
\begin{center}
\includegraphics[width=32pc]{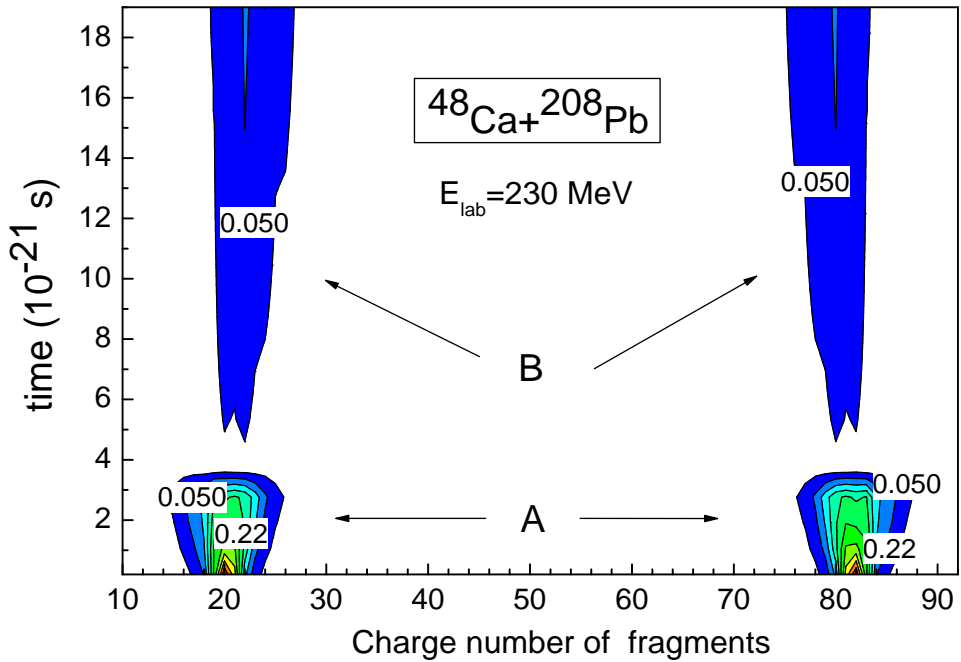}
\vspace*{-4.5cm}
\caption{\label{DICQfisCaPb} Time dependence of the charge distribution
of deep inelastic collision (A) and quasifission (B) products for the
$^{48}$Ca + $^{208}$Pb reaction.}
\end{center}
\end{figure}

We would like to stress the importance to distinguish the scattering processes,
with no or negligible mass drift, i.e. quasi- and deep inelastic processes,
from capture reactions that exhibit a small or large mass drift,
usually but not always going to the full mass symmetry configuration and leading to decay into two almost equally large fragments.

\begin{figure}[t] 
\begin{center}
\includegraphics[width=32pc]{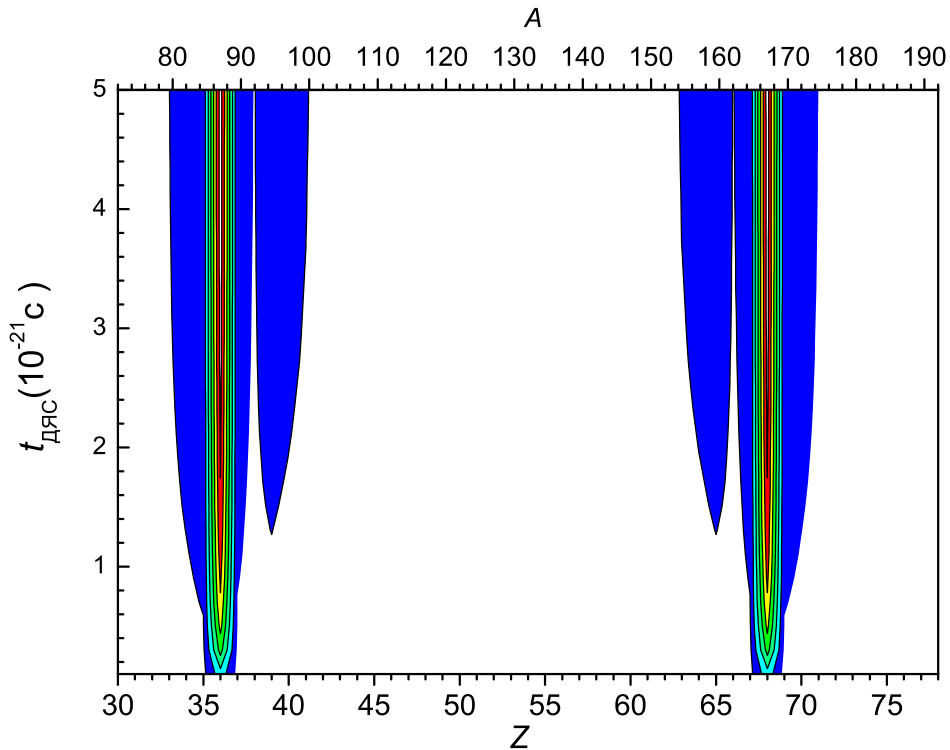}
\vspace*{-0.5cm}
\caption{\label{Zmax} Time dependence of the charge distribution
of deep inelastic collision and quasifission products for the
$^{86}$Kr + $^{166}$Er reaction.}
\end{center}
\end{figure}

The analysis of the mass (charge) distribution of the binary
reaction products showed in Refs. \cite{RudolfNPA330} and
\cite{FazioMPL2005,Prokhorova} shows that in collisions of nuclei
with  magic proton or/and neutron numbers at a charge or mass
drift  around Coulomb barrier energies  may be small in  capture
reactions too. In Ref. \cite{RudolfNPA330}, an averaged charge
number of the projectile-like products $^{86}$Kr+$^{166}$Er was
equal to the initial value $<Z_{\rm PLF}>=36$ up to the total
kinetic energy (TKE) loss of 180 MeV. Our theoretical results
presented in  Fig.\ref{Zmax} show that  deep inelastic and
quasifission reactions products are mixed having the same charge
number because the driving potential has a minimum at $Z=36$
corresponding to the projectile charge number. A similar
phenomenon was observed in the $^{48}$Ca+$^{208}$Pb reaction
\cite{Prokhorova}. The small drift from the initial charge number
of the projectile-like products does not mean that the deep
inelastic reactions occurred only. A maximum of the charge
distribution can be concentrated at the charge value corresponding
to the minimum of the potential energy surface. This phenomenon
was reproduced by the DNS model and our results of the charge and
mass distributions of the deep inelastic collisions (A) and
quasifission (B) fragments are presented in Fig.\ref{DICQfisCaPb}.

Understanding the mass drift is basic for the investigation of the
reaction mechanism. It is relevant to the deep inelastic processes
and to the capture processes. In capture reactions with heavy
nuclei, the direction of the mass drift could influence the
probability for the compound nucleus formation: a drift towards
asymmetry could favor true compound nucleus formation (with
subsequent evaporation of neutrons leading to the observed
evaporation residues or fission fragments), whereas a drift
towards symmetry could favor quasifission without compound nucleus
formation.

\section{Difference between capture and complete fusion for light and massive systems}

For light or medium-heavy systems, capture inside the Coulomb barrier leads to
complete fusion, so that the capture (or barrier-passing) cross section may be
equal to the complete fusion cross-section.  According to the DNS concept
 the hindrance to complete fusion is connected with the presence of the quasifission
process  as a competing channel which leads to formation of binary products without
formation of the compound nucleus \cite{GiarSHE,FazioJPSJ72}.
For the analysis and study of the hindrance to complete fusion we
calculate the potential energy surface for DNS formed during the interaction of the nuclei
as the sum of the reaction energy balance ($Q_{\rm gg}$) and the nucleus-nucleus
interaction $V(R)$:
\begin{equation}
U_{\rm driv}(Z_1,R,\ell)=Q_{\rm gg}+V(R),
\end{equation}
\begin{figure}[t]   
\begin{center}
\includegraphics[width=32pc]{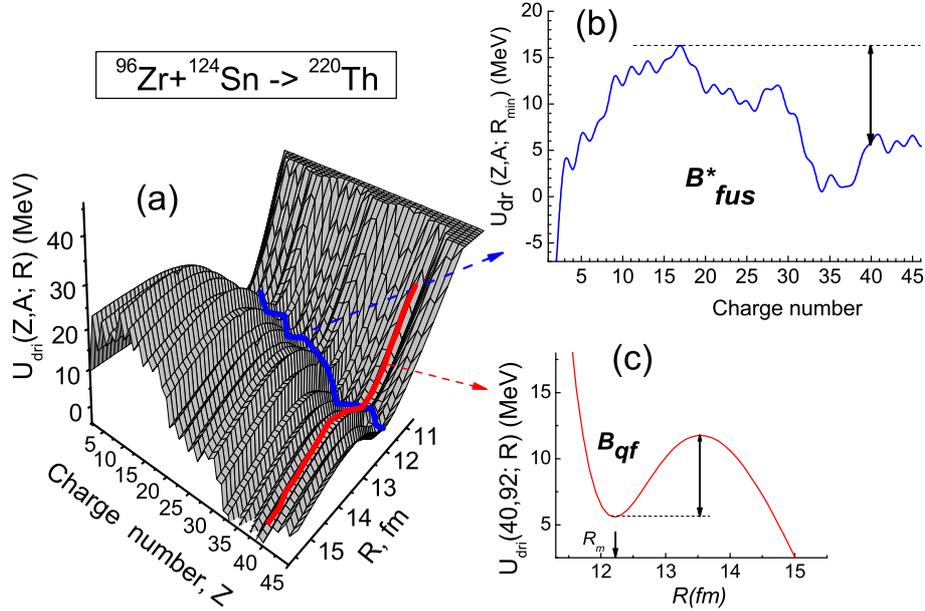}
\vspace*{-0.5cm}
\caption{\label{PESTh214} Potential energy surface calculated for the DNS
 formed in the $^{96}$Zr+$^{124}$Sn reaction (a); driving potential
 built by connection of minima of the potential wells for all values
 of charge asymmetry of DNS (b): $B^*_{\rm fus}$ is considered
 as the intrinsic fusion barrier for the transformation of the DNS from the
 configuration with the charge asymmetry $Z=40$;
 potential well with depth $B_{\rm qf}$
 which is used as the quasifission barrier (c). The minimum value of
 the potential well is at $R=R_m$.}
\end{center}
\end{figure}
where $Q_{\rm gg}=B_1+B_2-B_{\rm CN}$; $B_1$,  $B_2$ and $B_{CN}$ are the binding energies of the
constituents of the DNS and compound nucleus, respectively. The values of
 the binding energies
are obtained from \cite{MassAW95,Moeller}. The evolution of DNS
with given initial charge numbers of projectile- and
target-nucleus ($Z_P$ and $Z_T$) is determined by the potential
energy surface. As an example, the potential energy surface
calculated for the $^{96}$Zr+$^{124}$Sn reaction is presented in
Fig. \ref{PESTh214}(a). The approaching path of the projectile to
the target nucleus occurs along the $R$-axis which is the distance
between mass centers of colliding nuclei. Overcoming the Coulomb
barrier the system starts to lose its kinetic energy of relative
motion due to particle-hole excitation of nucleons and nucleon
exchange as it is shown in Fig. \ref{CapDIC}. In this Section and
further we consider capture reactions when a full momentum
transfer in heavy ion collisions occurs. The evolution of DNS
along charge asymmetry takes place along the valley on the
potential energy surface.  The driving potential $U_{\rm driv}$ is
the curve lying on the bottom of this valley. A presence of a
hindrance to complete fusion by multinucleon transfer can be found
by an estimation of the intrinsic fusion barrier  $B^*_{\rm fus}$
as the difference between a values of $U_{\rm driv}$ corresponding
to the initial charge asymmetry and its maximum value to the way
of complete fusion $B^*_{\rm fus}(Z_P)=U_{\rm driv}(Z_{\rm
max})-U_{\rm driv}(Z_P)$. For example, in Fig. \ref{PESTh214}(b),
$B^*_{\rm fus}(Z_P)$=10.2 MeV, because $U_{\rm driv}(17)=16$ and
$U_{\rm driv}(40)=5.8$ MeV. The quasifission barrier $B_{\rm qf}$
for the charge asymmetry $Z_P=40$ is equal to 5.68 MeV. The
excitation energy of the DNS $E^*_{\rm DNS}=E_{\rm
c.m.}-V(R_m)+\Delta Q_{\rm gg}$, as well as  $B^*_{\rm fus}$ and
$B_{\rm qf}$ determine a hindrance to complete fusion.

\begin{figure}[t]  
\begin{center}
\includegraphics[width=32pc]{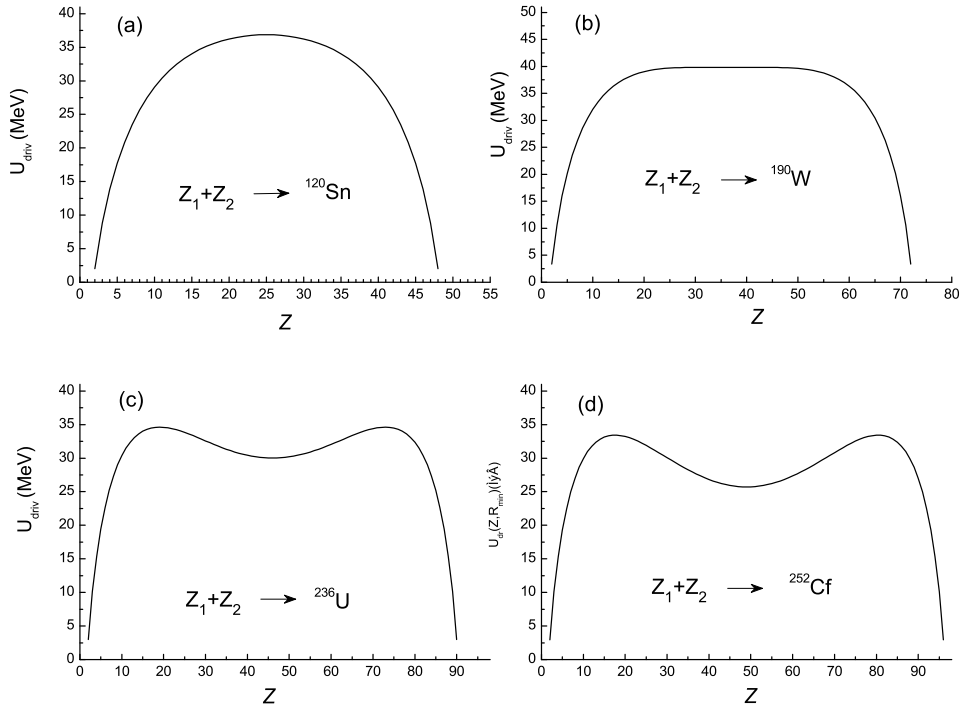}
\vspace*{-0.5cm}
\caption{\label{Driv4Nuclei} Driving potentials calculated for the DNS
leading to formation of the compound nuclei $^{120}$Sn (a), $^{190}$W (b),
 $^{236}$U (c) and $^{252}$Cf (d).}
\end{center}
\end{figure}

\begin{figure}[t] 
\begin{center}
\includegraphics[width=32pc]{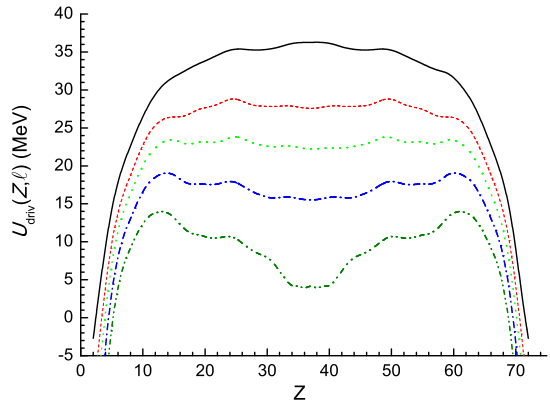}
\vspace*{-4.0cm}
\caption{\label{DrivL190W} The dependence of the driving potential
on the DNS angular momentum $\ell_{\rm DNS}$ for the reactions
leading to the CN $^{190}$W: $\ell=0$ (solid line), 40 (dashed line),
60 (dotted line), 80 (dot-dashed line), and 100 (double dot-dashed line).
}
\end{center}
\end{figure}

Taking into account the dependence of the hindrance to complete
fusion on the orbital angular momentum $\ell$
we calculate partial fusion cross section  as a function of the
 collision energy $E_{\rm c.m.}$ \cite{GiarSHE,FazioJPSJ72,Nasirov05}:
\begin{equation}
\label{Sigmafus}
\sigma_{\rm fus}^{(\ell)}(E_{\rm c.m.})=\sigma_{\rm cap}^{(\ell)}(E_{\rm c.m.})
P_{\rm CN}^{(\ell)}(E_{\rm c.m.}),
\end{equation}
$P_{\rm CN}^{(\ell)}$ is the fusion probability in competition
between complete fusion and quasifission. For light systems the
capture and complete fusion cross sections are equal because
$P_{\rm CN}^{(\ell)}=1$ for the values of $\ell$ for which the
intrinsic fusion barrier $B_{\rm fus}^*$ is equal to zero. To see
a dependence of the fusion probability on the total charge and
mass numbers of the colliding nuclei we compare the driving
potentials ($\ell=0$) which were calculated by the use of the
liquid-drop model to obtain the binding energies $B_1$, $B_2$, and
$B_{\rm CN}$ for the four reactions leading to the compound nuclei
$^{120}$Sn, $^{190}$W, $^{236}$U and $^{252}$Cf (see, Fig.
\ref{Driv4Nuclei}). One can say that for all reactions leading to
the CN $^{120}$Sn there is no hindrance to complete fusion because
the maximum value of $U_{\rm driv}$ is at $Z_1=Z_2$. Therefore,
there is no intrinsic fusion barrier, $B^*_{\rm fus}$=0. For the
reactions leading to the CN $^{190}$W it is seen that
approximately $B^*_{\rm fus}$=0, but there is the plateau at
$20<Z<50$ which can be transformed to a hollow at large values of
angular momentum. As a result we can observe a hindrance to
complete fusion for colliding energies $E_{\rm c.m.}$  near and
above the Coulomb barrier. The dependence of the driving potential
on the DNS angular momentum for the reactions leading to the CN
$^{190}$W is demonstrated in Fig. \ref{DrivL190W}. The increase of
$\ell$ leads to decrease the middle charge part ($(Z_1+Z_2)/2$
region) of the driving potential. Consequently, a hindrance to the
complete fusion appears for the charge symmetric reactions leading
to formation of $^{190}$W at large angular momentum. The increase
of the hindrance by increasing the angular momentum for the
massive systems was analyzed in Ref. \cite{FazioPRC2005}.

\section{Cold and hot fusion mechanisms in the DNS model}

 The dynamics of complete fusion and the role of the entrance  channel in the
  formation of heavy ion collision reactions are questionable or/and they have different
interpretation still now.
 For example, what mechanism of fusion makes the main contribution  to the formation
of the compound nucleus:  the increase of the radius of neck
between interacting nuclei or multinucleon transfer at relatively
restricted neck size?  Priority of the proposed mechanism is
determined by its possibility to explain different characteristics
of the observed physical quantities using the same dynamical
variables of the model. To show the advance of the DNS model we
demonstrated the peculiarities of the angular momentum
distribution of DNS  and compound nucleus which determine the
angular distribution of  reaction products, cross sections of
evaporation residue, fusion-fission and quasifission products.
 We are trying to improve this model to use it for the
separation of fusion-fission fragments from the quasifission and
fast fission products. It is well known that similarity of the
mass and energy distributions of above mentioned processes causes
difficulties at the analysis of the experimental data to establish
the contribution of the every reaction mechanism into measured
data. As well as this model can be applied to study distribution
of the excitation energy between different degrees of freedom, as
well as between reaction products.

In the DNS concept \cite{Antonenko}, the evaporation residue cross section at
the collision energy  $E_{\rm c.m.}$ is factorized as follows:
\begin{equation}
\label{ERcross}
\sigma_{\rm ER}(E_{\rm c.m.})=\sum_{\ell=0}^{\ell_{\rm f}}\sigma_{\rm fus}^{(\ell)}(E_{\rm c.m.})
W_{\rm sur}^{(\ell)}(E_{\rm c.m.}),
\end{equation}
where $\sigma_{\rm fus}^{(\ell)}$ is the partial cross section  of the complete
fusion of the projectile and target nuclei;
 $W_{\rm sur}^{(\ell)}$ is the survival probability against fission of the
heated and rotating nucleus at each step of the de-excitation
cascade by evaporation of neutrons, protons and light particles up
to the formation of the evaporation residue; $\ell_{\rm f}$ is the
value of CN angular momentum at which the fission barrier
disappears: $W_{\rm sur}^{(\ell)}(E)=0$ for $\ell > \ell_{\rm f}$.
The decrease of $W_{\rm sur}$ by increasing the excitation energy
is determined by the increase of rate of competition between
fission and emission of particles.

At the synthesis of the superheavy elements $Z>114$ the observed
cross sections of the evaporation residues are near or less than 1
pb. The smallness of $\sigma_{\rm fus}$ or/and $W_{\rm sur}$ in
formula (\ref{ERcross}) leads to small values of the measured
cross section $\sigma_{\rm ER}\sim 1$pb.
\begin{figure}[t] 
\vspace*{2.0cm}
\begin{center}
\includegraphics[width=32pc]{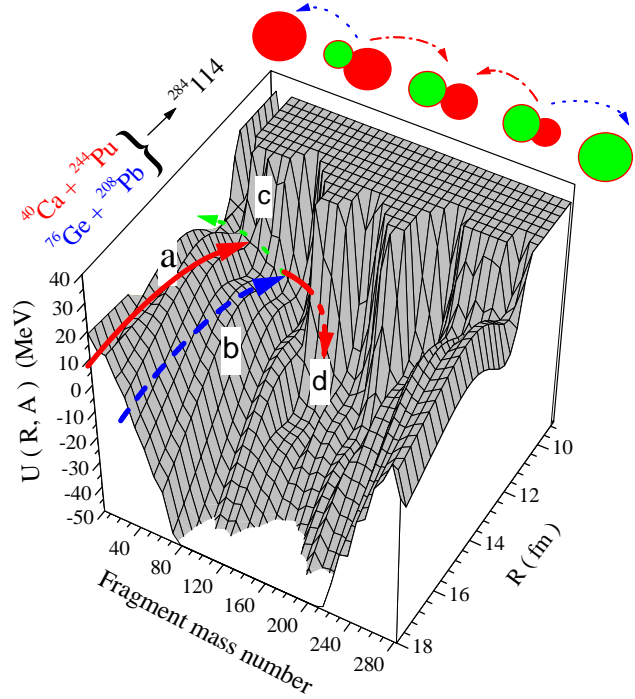}
\vspace*{-4.0cm} \caption{\label{DNSColdHot} An explanation of the
inevitable difference between excitation energies of the compound
nucleus which can be formed in the  ``hot'' ($^{40}$Ca+$^{244}$Pu
--(a)) and ``cold'' ($^{76}$Ge+$^{208}$Pb -- (b)) fusion reactions
by the entrance channel and peculiarities of the potential energy
surface. The arrows (c) and (d) show the complete fusion and
quasifission directions, respectively. }
\end{center}
\end{figure}

The experience of the synthesis of superheavy elements showed that
the used reactions can be separated into "cold" and "hot" fusion
reactions. In the "cold fusion" reactions the excitation energy
the formed compound nucleus $E^*_{\rm CN}$ is less than 20 MeV
\cite{SHofmann,Morita} while in "hot fusion" reactions $E^*_{\rm
CN}$ is more than 25 MeV \cite{Oganessian}. In the "cold fusion"
reactions the evaporation residue nuclei are formed after emission
1 or 2 neutrons from the heated and rotating compound nucleus. In
the "hot fusion" reactions the 3 or more neutron emission cascade
precedes the evaporation residue formation. The charge asymmetry
of the entrance channel determines the type of reaction: ``cold
fusion'' reactions with the $^{208}$Pb and $^{209}$Bi targets used
to synthesize superheavy elements Darmstadtium, Roentgenium and
Copernicium are more mass symmetric than the "hot fusion"
reactions with the $^{48}$Ca projectile on the actinide targets
used to synthesize the heaviest elements $Z=114$--118 in the
Flerov Laboratory of the Nuclear Reactions of Joint Institute for
Nuclear Research. In Fig. \ref{DNSColdHot}, the arrow "a" shows
the path of the entrance channel leading to the ``hot fusion''
reactions while the arrow "b" shows the path leading to ``cold
fusion'' reactions. Due to peculiarities of the landscape of the
potential energy surface for the massive system, the excitation
energies of DNS and compound nucleus for the symmetric charge
numbers (large reaction $Q_{\rm gg}$ values) are smaller in the
entrance channel with the $^{208}$Pb target-nucleus (around arrow
"b"):
 $E^*_{\rm CN}=E_{\rm c.m.}+Q_{\rm gg}$. It is seen from the shape of the
potential energy surface that the ``hot fusion'' reaction can not
be made ``colder'' because the capture of projectile by the
target-nucleus becomes impossible by decreasing the beam energy.
Note,  for this system leading to formation of $^{284}$114, the
``cold fusion'' reaction $^{76}$Ge+$^{208}$Pb can not be made
``warmer'' because if we increase the beam energy significantly
higher than the Coulomb barrier the system can not be captured due
to the small size of the potential well in the nucleus-nucleus
interaction and restricted value of the friction coefficient for
radial motion: we observe only deep inelastic collisions as in
Fig. \ref{CapDIC}(b). One can say that, for the ``cold fusion''
reaction, the increase of the beam energy enough higher than the
Coulomb barrier does not lead to an increase of the fusion cross
section as it is expected from the extra-extra push model
\cite{Blocki}. In the ``cold fusion'' reactions with $^{208}$Pb
and $^{209}$Bi targets an increase of the
 projectile charge number leads to the drastic hindrance to complete fusion
 as a result of the
increase of contributions of the quasifission and deep inelastic
collisions. This phenomenon explains the observed small cross
section in the synthesis of superheavy element $Z=113$  in the
$^{64}$Zn+$^{209}$Bi reaction \cite{Morita}.
\begin{table}[h]
\caption{\label{Table1} Comparison of the hindrance to formation
of compound nucleus in the ``cold'' and ``hot fusion'' reactions
used in the synthesis of superheavy elements.}
\begin{center}
\lineup
\begin{tabular}{*{7}{l}}
\br
`Cold fusion''\m&$\eta=\frac{A_2-A_1}{A_1+A_2}$&$\0 P_{\rm CN}$&``Hot fusion''&\m$\eta=\frac{A_2-A_1}{A_1+A_2}$& \m \0 $P_{\rm CN}$ \m\cr
 \m\0 reactions \m& &&\m \0 reactions &\m&\cr
\mr
$^{64}$Ni+$^{208}$Pb$ ^{\dag}$ \m &\0 0.529 &$1.4\cdot 10^{-7}$&$^{48}$Ca+$^{244}$Pu \m&\m\0 0.671\m&\m $4.96\cdot 10^{-2}\m$\cr
$^{64}$Ni+$^{209}$Bi$ ^{\dag}$ \m &\0 0.531 &$7.0\cdot 10^{-8}$&$^{48}$Ca+$^{243}$Am$ ^{\ddag}$ \m& \m\0 0.670\m &\m $5.02\cdot 10^{-2}\m$\cr
$^{70}$Zn+$^{208}$Pb$ ^{\dag}$ \m &\0 0.496 &$2.5\cdot 10^{-9}$&$^{48}$Ca+$^{248}$Cm$ ^{\ddag}$ \m& \m\0 0.676\m &\m $1.13\cdot 10^{-2}\m$\cr
$^{70}$Zn+$^{209}$Bi$ ^{\dag}$ \m &\0 0.498 &$5.2\cdot 10^{-10}$&$^{48}$Ca+$^{249}$Bk$ ^{\S}$ \m& \m\0 0.677\m &\m $5.06\cdot 10^{-3}\m$\cr
$^{76}$Ge+$^{208}$Pb$ ^{\dag}$ \m &\0 0.465 &$1.2\cdot 10^{-10}$&$^{48}$Ca+$^{249}$Cf$ ^{\ddag}$ \m& \m\0 0.677\m &\m $7.14\cdot 10^{-3}\m$\cr
\br
\end{tabular}
\end{center}
\vspace*{-0.3cm}
$^{\dag}$The estimations from Ref.\cite{GiarSHE} \\
$^{\ddag}$The estimations from Ref. \cite{FazioEPJ2004}\\
$^{\S}$The estimations from Ref. \cite{CaBk2011}\\
\end{table}

\begin{figure}[h]  
\begin{minipage}{18pc}
\includegraphics[width=20pc]{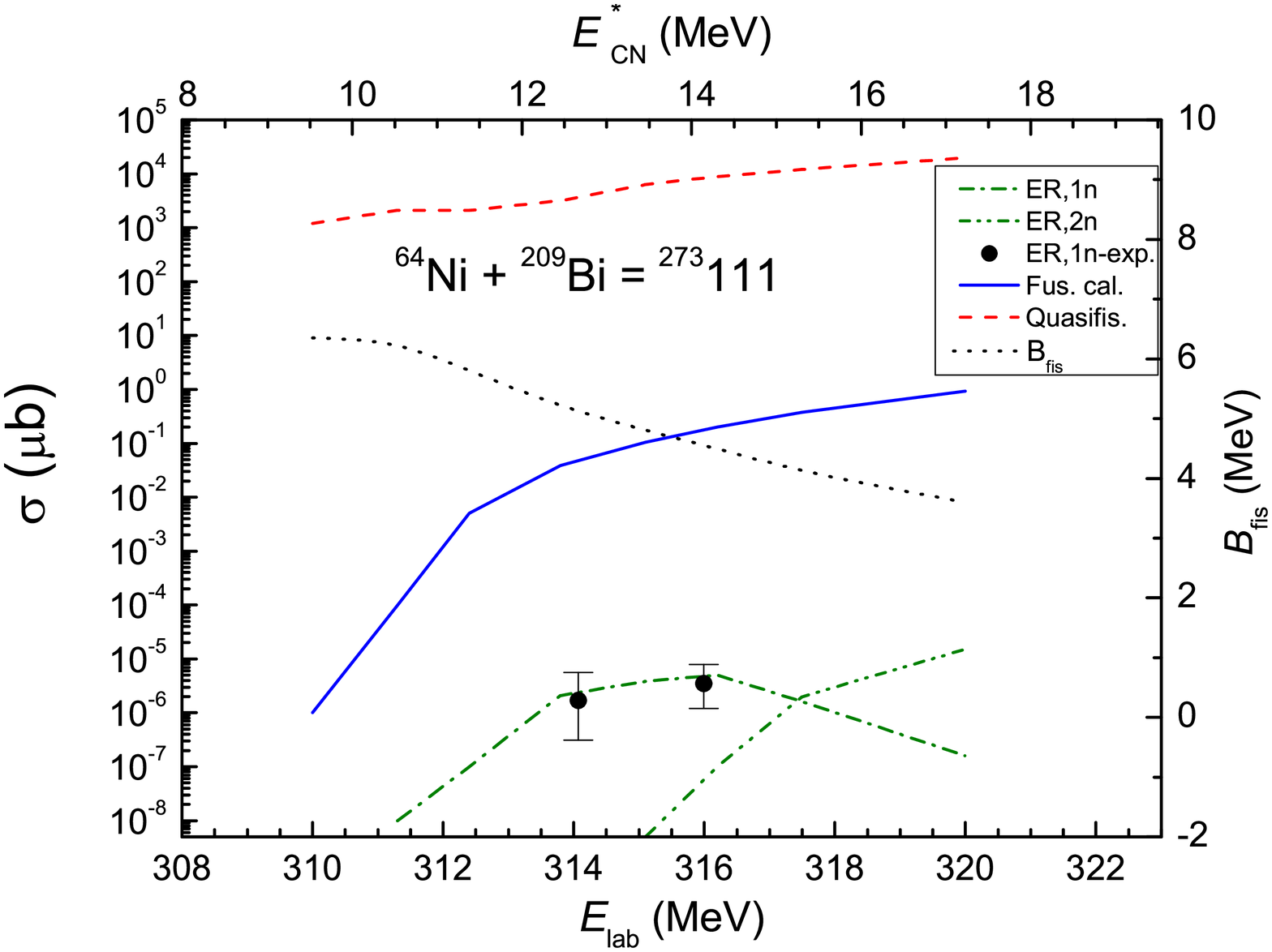}
\end{minipage}\hspace{0.5pc}%
\begin{minipage}{18pc}
\includegraphics[width=20pc]{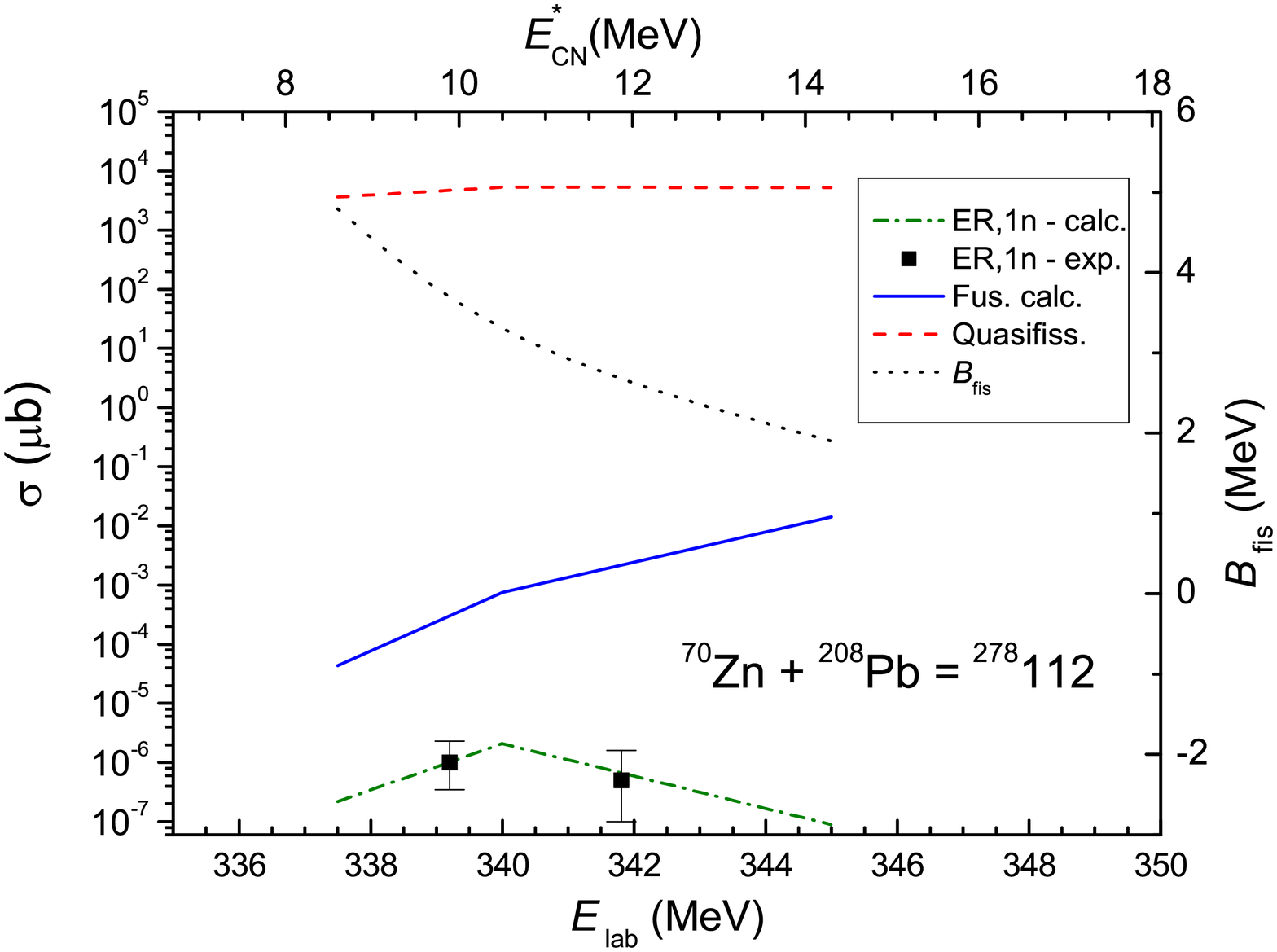}
\end{minipage}
\vspace*{-0.5 cm} \caption{\label{SHE111}Comparison of the ER
results (left scale) obtained by the DNS model (dot-dashed line)
with the experimental data for the synthesis of Roentgenium and
Copernicium in the $^{64}$Ni+$^{209}$Bi (left panel) and
$^{70}$Zn+$^{208}$Pb (right panel) reactions, respectively.
Quasifission and fusion excitation functions are shown by dashed
and solid lines, respectively. The dotted line is the fission
barrier (right scale).}
\end{figure}

 In Fig. \ref{SHE111}  we present a comparison of the
ER cross sections obtained by the DNS model (dot-dashed line) with the experimental
data for the synthesis of Roentgenium and Copernicium in the
 $^{64}$Ni+$^{209}$Bi and $^{70}$Zn+$^{208}$Pb reactions, respectively.
The dashed and solid lines in Fig. \ref{SHE111} show the
theoretical results for the quasifission and complete fusion
excitation functions. The ratio of the complete fusion to capture
excitation function $(P_{\rm CN})$ shows how strong is the
hindrance to complete fusion (see Eq. (\ref{Sigmafus})) due to the
very small values of fusion cross section in comparison to the
capture cross section. The event at $E^*_{\rm CN}=9$ MeV in the
$^{70}$Zn+$^{208}$Pb reactions (right panel) was described and
published in \cite{GiarSHE} while the second event at $E^*_{\rm
CN}=12$ MeV was measured after the appearance of the cited paper.
One can say that the second  event was predicted in
\cite{GiarSHE}.
 So, we conclude that the smallest cross section
 in the synthesis of Copernicium is caused mainly by the huge contribution
of the quasifission events that is inherent for the ``cold
fusion'' reactions. This conclusion was deduced from the
theoretical analysis which included realistic nuclear shell
effects in the calculation of the potential energy surface. The
dotted lines in Fig. \ref{SHE111} show the fission barrier of the
compound nucleus $B_{\rm fis}$ which is a function of its
excitation energy and angular momentum.

The ``hot fusion'' reactions were favorable for the synthesis of
the superheavy elements $Z>$112:  new superheavy elements
$Z=$114--118 have been obtained at the Flerov Laboratory of
Nuclear Reactions of JINR (Dubna, Russia) \cite{Oganessian} during
the last decade. The ER cross sections in the synthesis of
superheavy elements $Z=114$ and $Z=116$ were confirmed in the
recent experiments performed in the Lawrence Berkeley Laboratory
\cite{Duellmann} and GSI (Darmstadt) \cite{Heinz}, respectively.
The main reason allowing the experimentalists to succeed in the
``hot fusion'' reactions or the synthesis of the superheavy
elements $Z=$114--118 is smallness of the intrinsic fusion barrier
$B^*_{\rm fus}$ which decreases the hindrance to complete fusion.
But large values of the excitation energy and angular momentum of
the compound nucleus formed in the ``hot fusion'' reactions
decrease the survival probability against fission. Therefore, the
cross sections of evaporation residues in synthesis of superheavy
elements decreases from 3--4 picobarns for $Z=114$ up to 0.5 pb
for $Z=118$. In Table 1 we compare the hindrance for the formation
of the compound nucleus in the ``cold'' and ``hot fusion''
reactions which were used in the  synthesis of superheavy
elements. The difference in the hindrances presented by $P_{\rm
CN}=\sigma_{\rm fus}/(\sigma_{\rm fus}+\sigma_{\rm qfis}$)
 in both types of reactions is evidently seen:
 i) the hindrance to complete fusion is
stronger for the ``cold fusion'' reactions in comparison with
``hot fusion'' reactions; ii) the hindrance  to ``cold fusion''
reactions increases drastically by increasing of the projectile
charge number while it changes slowly in ``hot fusion'' reactions
by increasing the target charge number. This is explained by a
strong change of the potential energy surface around the charge
asymmetry $Z=82 (A=208)$ and it has more sloping shape in the
region $Z=20 (A=48)$ (see Fig. \ref{DNSColdHot}).

\section{Mixing of mass-angle distributions of the quasifission and
fusion-fission products}

The mass (charge) and angular distributions of the reaction products in
the heavy ion collisions are the main indicator  used to make conclusions
 about the reaction
mechanism. According to the established opinion the products with
masses close to the ones of projectile and target nuclei are
considered as products formed in deep inelastic reactions; the
products with the symmetric mass distributions are considered as
products of the fusion-fission reactions, {\it i.e.} fission
products of the heated and rotating compound nucleus formed at
complete fusion; the products having intermediate masses between
projectile-like and fusion-fission products are considered as
quasifission products. The last kind of products appear when
proton or neutron numbers in the heavy or light fragment are close
to the magic numbers 28, 50, 82 or 126 \cite{Itkis734,Knyazheva}.
So, the symmetric mass distributions from capture reactions can
either be the result of a compound nucleus formation followed by
fission, or - if the criteria for compound decay are violated -
they are ascribed to the  quasifission and fast fission  channels.
\begin{figure}[t]  
\begin{center}
\includegraphics[width=32pc]{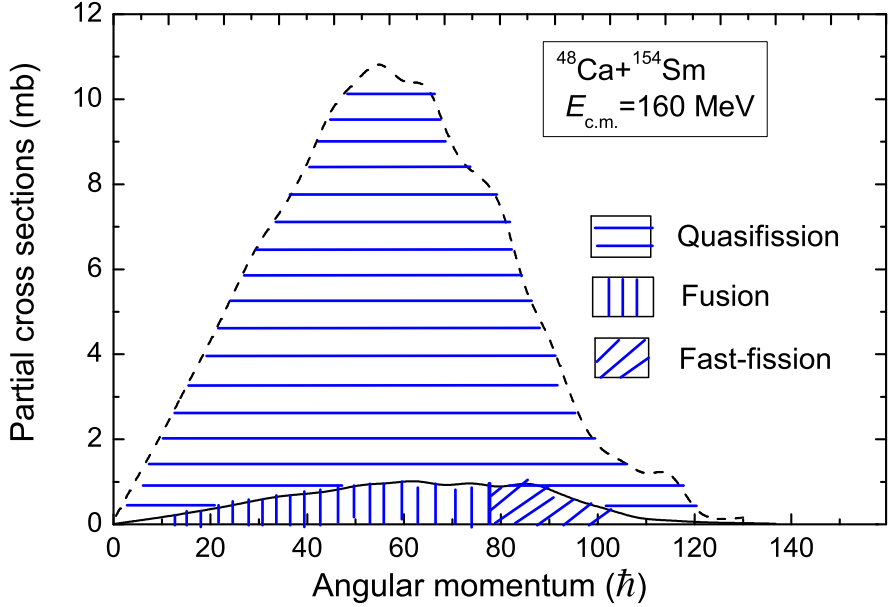}
\vspace*{-0.85cm}
\caption{\label{PartCa48Sm154} Partial cross sections (angular momentum distributions)
of quasifission (area filled by the horizontal lines), fast fission (area filled by the
skew-eyed lines) and complete fusion (area filled by the
vertical lines) events calculated by the DNS model
for the $^{48}$Ca+$^{154}$Sm reaction.}
\end{center}
\end{figure}

We should remind that quasifission products are formed in the
decay of the DNS without reaching the stage of the compound
nucleus or one can say without reaching the saddle point to arrive
to the compact shape. Quasifission means that the TKE of binary
fragments is close to that of fusion-fission products or Viola
systematics.

The experimental methods used to estimate the fusion
probability depend on the unambiguity of identification of the
complete fusion reaction products among the quasifission products. The
difficulties arise when the mass (charge) and angular
distributions of the quasifission and fusion-fission fragments
strongly overlap depending on the reaction dynamics. As a result,
the complete fusion cross sections may be overestimated:
\begin{equation}
\label{totfus} \sigma_{\rm fus}=\sigma_{\rm ffis}+\sigma_{\rm ER},
\end{equation}
where $\sigma_{\rm ER}$ is measured by good accuracy while
$\sigma_{\rm ffis}$ may include contribution of the quasifission
and fast fission products which are formed in the decay of the DNS
and in the  decay of the deformed mononucleus-no compound nucleus.
We remind the difference between quasifission and fast fission
processes:

-- Quasifission is the decay of DNS which is formed in the
capture--full momentum
 transfer reactions. The angular momentum distribution for the quasifission events
 can extend from $\ell=0$ up to $\ell=\ell_{\rm d}$ where  $\ell=\ell_{\rm d}$
the maximum value of the DNS angular momentum (see Fig. \ref{PartCa48Sm154},
 area filled by the horizontal lines).
The mass (charge) distribution of quasifission products may be in the wide range
 from masses (charges) of projectile- and target-like fragments up to
symmetric masses mixing with the fusion-fission products.

-- Fast fission is the decay of the mononucleus which is a
non-equilibrated
   system survived against quasifission  but unstable to be formed as
   compound nucleus due to the absence of a fission barrier caused by its
   fast rotational velocity \cite{Sierk}. Therefore, the partial
  cross section of fast fission is only populated  at  large values of the
   angular momentum  $\ell$ (see Fig.\ref{PartCa48Sm154},
  the area filled by skew-eyed lines).

\begin{figure}[t] 
\begin{center}
\includegraphics[width=32pc]{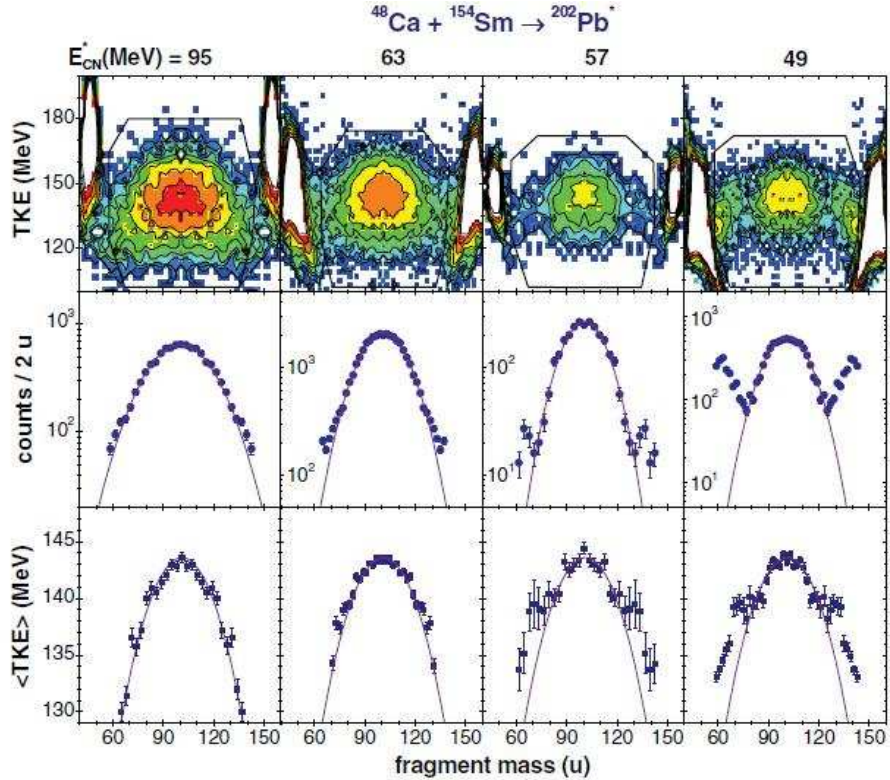}
\vspace*{-4.0cm} \caption{\label{QfissCaSm} Two-dimensional
TKE–mass matrices (upper panels), yields of fragments and their
TKE as a function of the fragment mass (middle and bottom panels,
respectively) in the $^{48}$Ca+$^{154}$Sm reaction at  different
$E^*_{\rm CN}$ excitation energies (designated above the upper
panels). Solid lines in the middle and bottom panels are Gaussian
and parabola fits
 to the mass and TKE distributions, respectively. This figure was
 copied from Ref. \cite{Knyazheva}.}
\end{center}
\end{figure}

\begin{figure}[t] 
\begin{center}
\includegraphics[width=35pc]{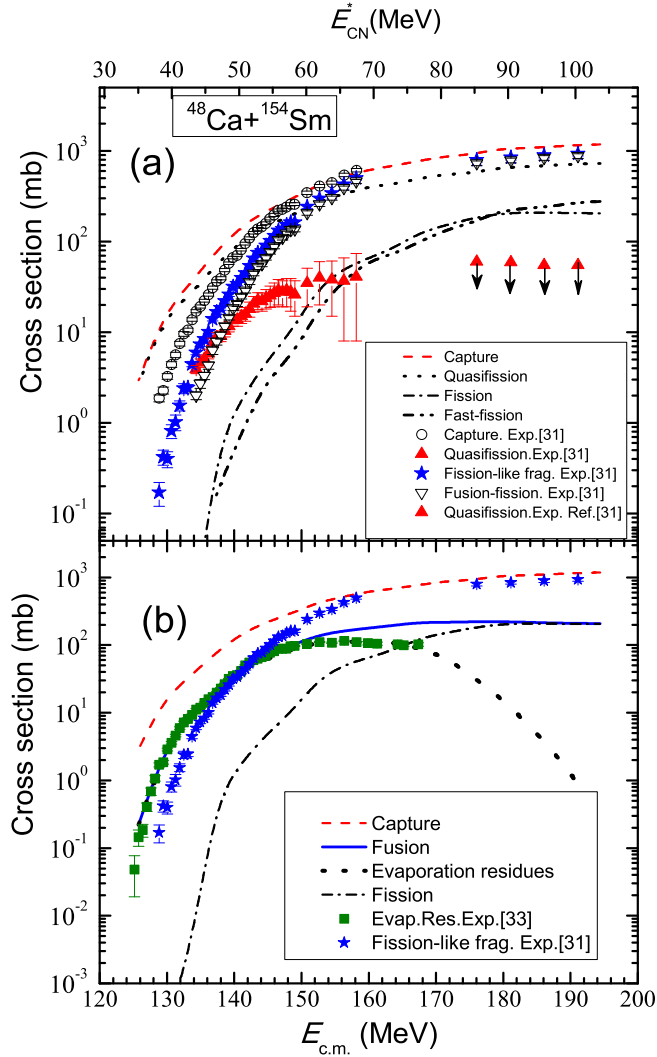}
\vspace*{-7.25cm}
\caption{\label{CompCa154Sm} Comparison of the
results of this work by the DNS model for the capture, complete
fusion, quasifission, fast fission and evaporation residue cross
sections with the measured data of the fusion-fission and
quasifission given in Ref. \cite{Knyazheva} (panel (a)) and with
data of the evaporation residues obtained from Ref.
\cite{Stefanini} (panel (b)) for the $^{48}$Ca+$^{154}$Sm
reaction.}
\end{center}
\end{figure}

We tried to analyze  the reasons for the lack or disappearance of
the quasifission feature in the experimental data of the
$^{48}$Ca+$^{144}$Sm and $^{48}$Ca+$^{154}$Sm reactions presented
in the paper \cite{Knyazheva} as the main conclusion of study. The
authors established the fusion suppression and the presence of
quasifission for the reactions with the deformed $^{154}$Sm target
at beam energies near and below the Coulomb barrier. The ratio of
the quasifission fragments with masses in the range $55<A<145$ to
the total mass distribution of fission fragments decreases, with
respect to the contribution of the symmetric compound
nucleus-fission, as the $^{48}$Ca projectile energy increases (see
Fig. \ref{QfissCaSm}). The authors did not consider the
possibility of  quasifission product yield with masses outside the
range $55<A<145$. The quasifission products are formed in the
ranges $A<55$ and $A>145$ too. But those can be mixed with the
products of deep inelastic reactions as in the
$^{48}$Ca+$^{208}$Pb reaction
 which was discussed in Section \ref{Section2}. This is a reason why
 our theoretical results of the capture cross section (long dashed line)
overestimated the experimental data (open circles) presented in Ref. \cite{Knyazheva}
at the low energies (see  Fig. \ref{CompCa154Sm}(a)).

The origin of the measured fission-like fragments (stars) at large
bombarding energies is explained by the sum of the quasifission
(short dashed line), fusion-fission (dash dotted line) and fast
fission (dash-double-dotted line) fragments (see Fig. neutrons and
gamma-quanta \ref{CompCa154Sm}). The cross section of the fast
fission channel increases by increasing the bombarding energy due
to the increase of the angular momentum of mononucleus.
 At low energies the contribution of the fusion-fission to the yield
of binary fragments is small in comparison with the quasifission
contribution. The small calculated fusion-fission cross
section is explained by the large fission barrier ($B_{\rm
f}$=12.33 MeV) for the $^{202}$Pb nucleus according to the
rotating finite range model by A. J. Sierk \cite{Sierk} and by the
additional barrier $B_{\rm f}^{(\rm micr)}=-\delta
W=-(\delta W_{\rm saddle-point}-\delta W_{\rm gs})\cong 8.22$ MeV
caused by the nuclear shell structure. We conclude that the
experimental fusion-fission data obtained at low energy
collisions contain a huge contribution of
quasifission fragments with masses $A>83$ which show an isotropic
distribution as presented in Ref. \cite{Knyazheva}. This is not a
new phenomenon and it was discussed as a result of theoretical
studies, for example, in our previous papers
\cite{NasirovRauis,FazioMess}  and in Ref. \cite{AritomoNP744}.
 The experimental results confirming this  conclusion
appeared recently in Ref.  \cite{HindePRL101,KnyazhevaPNL}. The
contribution of the mass symmetric products mixed with the
fusion-fission products with similar masses increases the
ambiguity of  estimations of fusion cross section by formula
(\ref{totfus}). This is the  reason for the difference between our
theoretical results and the extracted ones from the experimental
data for quasifission.
\begin{figure}[t] 
\begin{center}
\includegraphics[width=32pc]{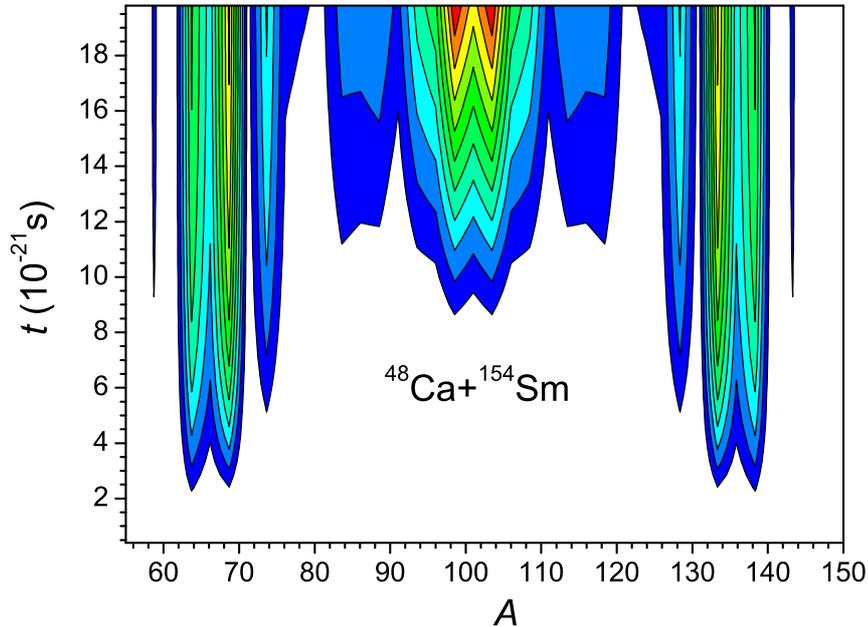}
\vspace*{-0.75cm}
\caption{\label{QfisMasDis} Mass distribution of quasifission
products formed in the $^{48}$Ca+$^{154}$Sm
reaction as a function of time.}
\end{center}
\end{figure}

 At the large energy $E_{\rm c.m.}$=154 MeV ($E^*_{\rm CN}$=63 MeV) the
experimental values of the quasifission cross section are much
lower than those of the fusion-fission cross section
\cite{Knyazheva}. According to our theoretical result a sufficient
part of the quasifission fragments shows the behaviour of
fusion-fission fragments: the mass distribution can reach the mass
symmetric region and their angular distribution can be isotropic
due to the possibility that the dinuclear system rotates by large
angles for large values of its angular momentum \cite{Varna10}.
The authors of Ref. \cite{Knyazheva} did not exclude such a
behaviour of the quasifission fragments. It is difficult to
separate  the quasifission fragments from the fusion-fission
fragments when both, their mass and angle distributions, overlap
in the region of symmetric masses. So, ignoring the quasifission
products mixed with deep inelastic collisions and considering all
the mass symmetric products as fusion-fission products give the
reasons for the  difference between the theoretical (dashed line
in Fig. \ref{CompCa154Sm}(a)) and experimental (up filled triangle
in Fig. \ref{CompCa154Sm}(a)) values of quasifission cross
sections.

At low energies the projectile-like quasifission fragments with
$A<70$ give a large contribution to the cross section for the
considered $^{48}$Ca+$^{154}$Sm reaction since the excitation
energy of the DNS is too small to shift the maximum of the mass
distribution to more mass symmetric configurations of the DNS. The
observed quasifission features at low energies are connected with
the peculiarities of the shell structure of the interacting
nuclei. The increase in the beam energy leads to a decrease of the
shell effects and the yield of the quasifission fragments near the
asymmetric shoulders decreases because the main contribution of
quasifission moves to the mass symmetric range. As it is seen from
Fig. \ref{QfisMasDis},  the yield of products with the masses in
the range $85<A<125$ appears at times $t>8.5\cdot10^{-21}$s. The
evaporation residue and fusion-fission excitation function were
calculated by the advanced statistical model
\cite{DArrigo92,DArrigo94}. In this model, the partial fusion
cross sections obtained by the DNS model were used as input data.
\begin{figure}[h]   
\vspace{-5.5pc}%
\begin{minipage}{20pc}
\hspace{-4pc}%
\includegraphics[width=28pc]{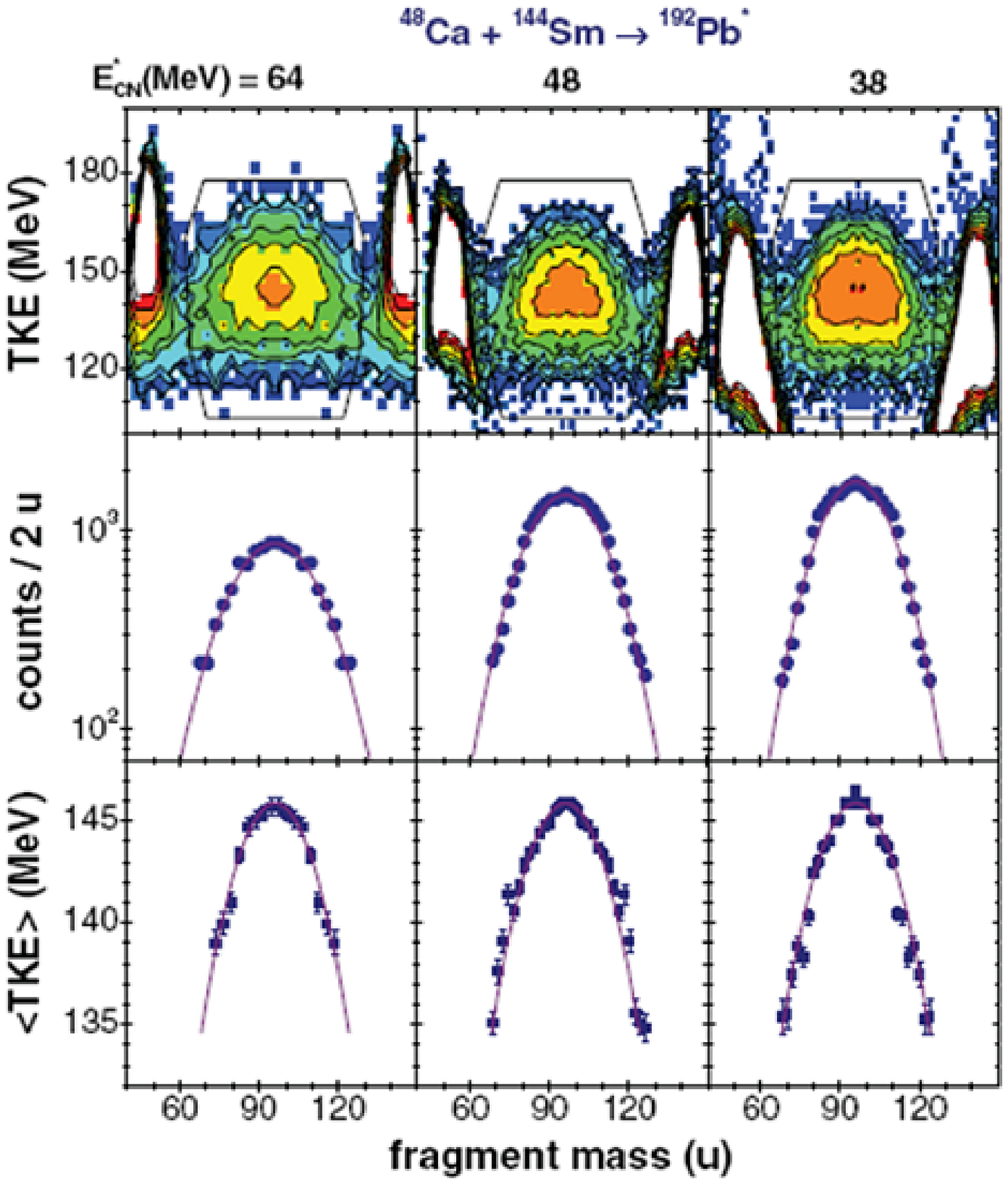}
\vspace{-8.8pc}%
\caption{\label{TKEMas48Ca144Sm}Two-dimensional TKE–mass matrices
(upper panels), yields of fragments and their TKE as a function of
the fragment mass (middle and bottom panels, respectively) in the
$^{48}$Ca+$^{144}$Sm reaction at the different $E^*_{\rm CN}$
excitation energies (designated above the upper panels). Solid
lines in the middle and bottom panels are Gaussian and parabola
fits
 to the mass and TKE distributions, respectively. This figure was
 copied from Ref. \cite{Knyazheva}.}
\end{minipage}
\hspace{1pc}%
\begin{minipage}{16pc}     
\hspace{-4pc}%
\vspace{-8.06pc}%
\includegraphics[width=24pc]{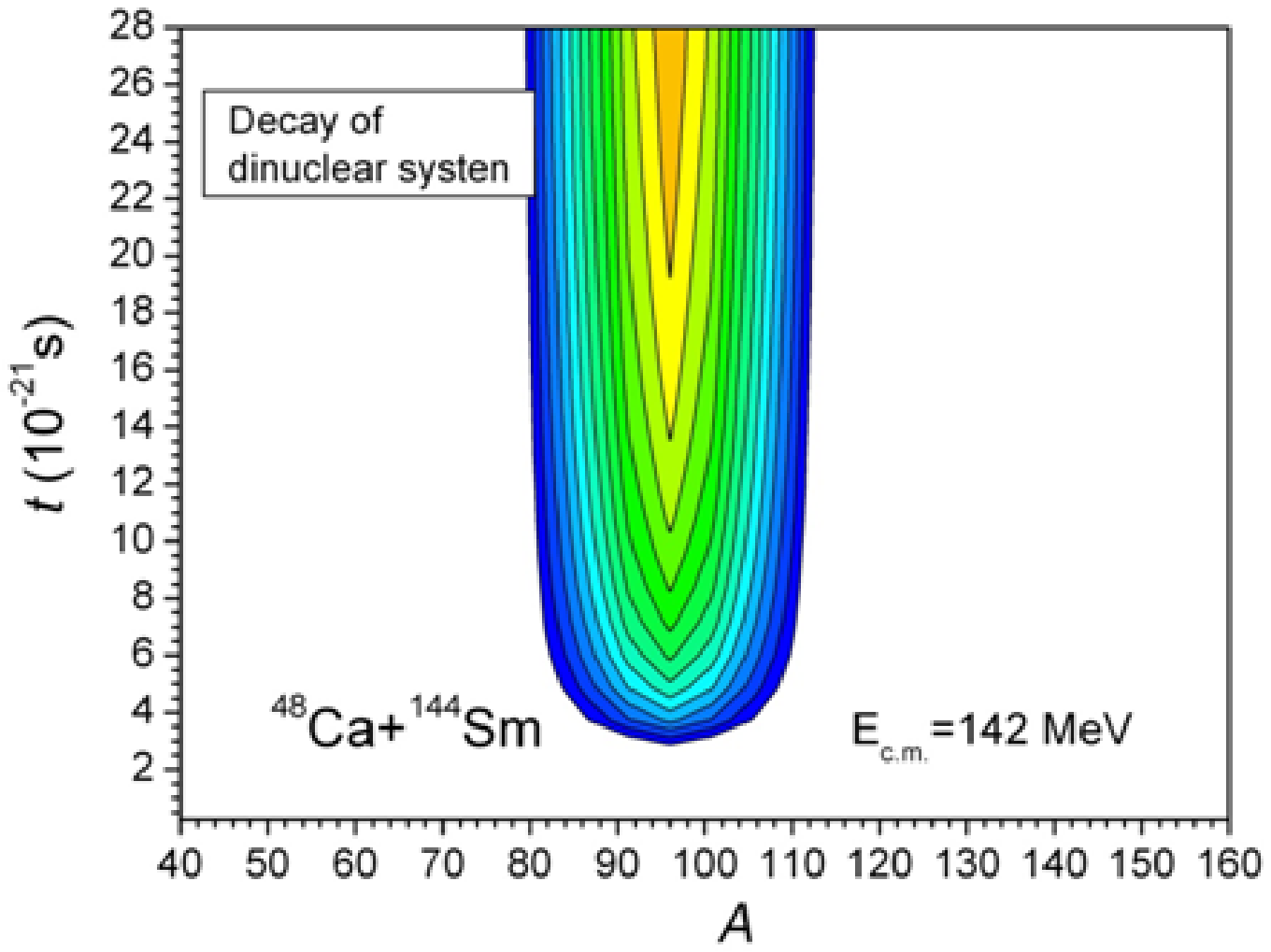}
\caption{\label{Yield48Ca144Sm}The mass distribution of the
quasifission products formed in the $^{48}$Ca+$^{144}$Sm reaction
as a function of time.}
\end{minipage}
\end{figure}

Due to large fission barriers of the compound nucleus $^{202}$Pb
the fission probability is very small and the yield of the
evaporation residues after neutron emission cascade dominates.
Therefore, our theoretical fission excitation function (dot-dashed
line) presented in  Fig. \ref{CompCa154Sm}(a) is sufficiently
lower than the experimental data (down open triangles). As we have
discussed above the latter data included a contribution of the
quasifission events. It is important to note that the evaporation
residue excitation function (thick dotted line in Fig.
\ref{CompCa154Sm}(b)) is in good agreement with the experimental
data (solid squares in Fig. \ref{CompCa154Sm}(b)) which is the
unambiguous physical quantity obtained from Ref. \cite{Stefanini}.
This fact confirms the correctness of our partial fusion cross
sections.

\section{Conclusions}

 The difficulties in the identification of the quasifission
products are connected with absence of the appropriate knowledge
about the reaction mechanism and a complete analysis of the yields
of quasifission and accompanying particles. The assumption about
the  possibility to observe the yield of quasifission products
only in the mass range in the middle between projectile-like and
fusion-fission products is not completely true: if there is no
sufficient yield of reaction products in this
 intermediate region it does not mean that there is no yield of the
 quasifission products. For example, the authors of Ref. \cite{Knyazheva}
 came to the same conclusion about the $^{48}$Ca+$^{144}$Sm reaction
 where they did not observe the characteristic peak of the
 quasifission products in the mass distribution (see Fig. \ref{TKEMas48Ca144Sm})
 Our theoretical results
 of the charge and mass distributions show that there are
  two reasons of a lack of the usual quasifission peak:
(i) one part of the mass distribution of the quasifission fragments is
 in the mass range of projectile-like products $48 < A < 60$;
 (ii) another part of the quasifission fragments is mixed
with the fusion-fission fragments and has similar isotropic
distributions.  The isotope $^{144}$Sm is a magic nucleus with the
neutron number $N$=82. Therefore, the concentration of the
asymmetric mode of the quasifission fragments in the mass range
$48 < A < 60$ is explained by the effect of the shell structure of
the double magic projectile-nucleus $^{48}$Ca and magic
target-nucleus $^{144}$Sm on the mass distribution of the reaction
fragments. As a result, the mass distributions of the products of
deep-inelastic collisions and asymmetric quasifission overlap in
this mass range. This case is similar to the $^{48}$Ca+$^{208}$Pb
reaction where the presence of the quasifission feature is
doubtful.
 But our investigation shows that due to the collision of the double magic
 $^{48}$Ca and $^{208}$Pb nuclei the mass distribution of the
quasifission fragments is concentrated around the initial masses
(see Fig. \ref{CapDIC}) because the potential energy surface has a
local minimum in this region. In Fig. \ref{Yield48Ca144Sm}, we
present the time dependence of the mass distribution of
quasifission products of the $^{48}$Ca+$^{144}$Sm reaction. One
can see that the mass numbers of the quasifission products are
concentrated in the mass range $80<A<110$ which overlaps
completely with the mass range of fusion-fission products. In
different from the $^{48}$Ca+$^{154}$Sm reaction, in the
$^{48}$Ca+$^{144}$Sm reaction no yield of the quasifission
products were found in the intermediate mass range where
quasifission is usually observed.

The lack of quasifission events in the experimental studies of the
$^{48}$Ca+$^{144}$Sm reaction, {\it i.e.} disappearance of
quasifission events by increasing the beam energy, is connected
with the measurement and analysis of the experimental data. More
advanced experiments at different beam energies and reaction
charge asymmetries
 must be performed by measuring neutrons and gamma-quanta in coincidence
 with the fission-like products. The results of such experiments should be analyzed
 by studying  the
different correlation functions of the observed quantities to
distinguish quasifission features in the case that the mass-angle
distributions of the quasifission and fusion-fission fragments
strongly overlap in the mass symmetric region. We should have
information about the angular momentum and excitation energy of
the system going to fission (compound nucleus or DNS) additionally
to the mass, angle and kinetic energy distributions of the fission
fragments and the accompanying neutrons. The theoretical models
must be enough perfect and they should have a small number of free
parameters to make unambiguous conclusions about the reaction
mechanism after a good description of the observed experimental
data which are interpreted without additional assumptions.

\end{document}